\documentclass{elsPLF-Note1}
 
\usepackage{amssymb}

\usepackage[english,francais]{babel}

\newtheorem{theorem}{Theorem}

\newtheorem{e-proposition}[theorem]{Proposition}

\newtheorem{e-definition}[theorem]{Definition\rm}

\renewcommand{\theequation}{\arabic{equation}}
\newcommand \Mcalb {\overline{\mathcal M}} 

\newcommand \auth {\textsc}

\newcommand \Mcal {\mathcal M}

\newcommand \Ker {\mbox{Ker}} 
\newcommand \dime {\mbox{dim }} 
 
\newcommand \Cotton  C
\newcommand \Kot {\mathcal K}

 \newcommand \TT T

\newcommand \Lcal {\mathcal L}

\newcommand \be         {\begin{equation}}           
\newcommand \ee         {\end{equation}}

\newcommand \RR         {\mathbb{R}}

\newcommand \del        \partial
\newcommand \eps   \varepsilon 
\newcommand \lam \lambda 

\newcommand \Rc {\textit{Rc}} 
\newcommand \Hcal {\mathcal H}
\newcommand \gbf {\mathbf g} 
 
\newcommand \Tbf {\mathbf T} 
 
\newcommand \Rbf {\mathbf R} 
\newcommand \Gbf {\mathbf G} 
 
\newcommand \Nbf {\mathbf N} 
 
\newcommand \la \langle 
\newcommand \ra \rangle

\def\og{\leavevmode\raise.3ex\hbox{$\scriptscriptstyle\langle\!\langle$~}}
\def\fg{\leavevmode\raise.3ex\hbox{~$\!\scriptscriptstyle\,\rangle\!\rangle$}}

\journal{Notes Comptes Rendus Acad. Sc. Paris (2010)}
\begin{document} 
\centerline{}
\begin{frontmatter}
 
\selectlanguage{english}
\title{Uniqueness of Kottler spacetime 
and Besse conjecture}

\selectlanguage{english}
\author[authorlabel1]{Philippe G. LeFloch} and 
\ead{pgLeFloch@gmail.com}
\author[authorlabel2]{Luc Rozoy}
\ead{Luc.Rozoy@ujf-grenoble.fr}

\address[authorlabel1]{Laboratoire Jacques-Louis Lions \& Centre National de la Recherche Scientifique, 
\\
Universit\'e Pierre et Marie Curie (Paris 6), 4 Place Jussieu, 75252 Paris, France.
\\ 
Blog: {\sl http://philippelefloch.wordpress.com.}}
\address[authorlabel2]{Institut Fourier, Universit\'e de Grenoble, 38402 Saint-Martin d'H\`eres, France.}

\medskip
\begin{center}
\end{center}

\begin{abstract}
\selectlanguage{english}
We establish a black hole uniqueness theorem for Schwarzschild-de~Sitter spacetime,   
also called Kottler spacetime, which 
satisfies 
Einstein's field equations of general relativity with positive cosmological constant. 
Our result concerns the class of static vacuum spacetimes with compact spacelike slices
and regular maximal level set of the lapse function.  
We provide a characterization of the interior domain of communication 
of the Kottler spacetimes, 
which surrounds an inner horizon 
and is surrounded by a cosmological horizon.  The proposed proof combines  
arguments from the theory of partial differential equations
and differential geometry,  
and is centered on a detailed study of a possibly singular foliation.
We also apply our technique in the Riemannian setting, and establish 
the validity of the so-called Besse conjecture.


\vskip 0.5\baselineskip 

\end{abstract}
\end{frontmatter}


\selectlanguage{english} 
\section{Introduction} 
Static vacuum spacetimes satisfying the Einstein field equations
 play a central role in general relativity, 
since such spacetimes are expected to represent a final state of the evolution of matter under self-gravitating forces. 
Several classical results show that, under certain physical conditions, a very limited number of such spacetimes exists. 
We are interested here in the class of spatially compact spacetimes  
with positive cosmological constant, which 
is not covered by the mathematical techniques available in the literature   
and, therefore, we establish here a new black hole uniqueness theorem. Our proof
overcomes several conceptual and technical
difficulties, as explained below. 

By definition, a {\em static spacetime with maximal compact spacelike slices} (of class $W^{2,2}(\Mcalb)$) 
is a time-oriented, $(3+1)$-dimensional Lorentzian manifold $\Nbf$ 
with global topology $\Nbf \simeq \RR \times \Mcal$ and Lorentzian metric 
$\gbf = - f^2 \, dt^2 + g$, 
where 
$t$ is a coordinate on $\RR$ increasing toward the future,   
$\Mcal$ is a connected, orientable, smooth topological $3$-manifold with smooth boundary $\del \Mcal$
such that $\Mcalb := \Mcal \cup \del \Mcal$ is compact and is  
endowed with a $t$-independent Riemannian metric $g$ of class $W^{2,2}(\Mcal)$, 
and $f: \Mcal \to (0, +\infty)$ belongs to the Sobolev space $W^{2,2}(\Mcal)$ and vanishes at the boundary.

The assumed regularity means that, in an atlas of local coordinates, 
the metric coefficients admit derivatives up to second-order that are squared-integrable. 
In this context, $f$ is referred to as the {\sl lapse function}, and 
the vector field $\Tbf := \del/\del t$ is a future-oriented, timelike Killing field: 
$$
\Lcal_\Tbf \gbf = 0, \qquad 
\gbf(\Tbf,\Tbf) < 0. 
$$
By definition, the hypersurfaces $t=$const. are orthogonal to $\Tbf$,
and
the spacetime  $\Nbf$ is foliated by compact spacelike slices with boundary. The lapse function 
is positive in $\Mcal$ and vanishes on $\del \Mcal$, so that 
the zero-level set of $f$ 
$$
\Hcal := \big\{ f=0 \big\},
$$
referred to as the {\sl horizon}, coincides with the boundary of the slices 
$
\Hcal = \del \Mcal 
$
(which need not be connected).

In addition, we impose that $\Nbf$ satisfies Einstein's vacuum equations with positive cosmological constant $\Lambda>0$, that is, $\Gbf_{\mu\nu} + \Lambda \, \gbf_{\mu\nu} = 0$, where 
$\Gbf_{\mu\nu} := \Rbf_{\mu\nu} - (\Rbf/2) \gbf_{\mu\nu}$ denotes Einstein's curvature tensor
(in dimensions $3+1$), $\Rbf_{\mu\nu}$ the Ricci curvature, 
and $\Rbf$ the scalar curvature, respectively. 
In other words, we impose 
$$
\Rbf_{\mu\nu} = \Lambda \, \gbf_{\mu\nu}. 
$$

Such a spacetime was discovered 
by Kottler \cite{Kottler}, and its most relevant part for us is the ``interior domain of communication'', 
defined as follows. 
Given $m, \Lambda>0$ satisfying $(3m)^2 \Lambda \in (0, 1)$, 
the {\em interior domain of the Kottler spacetime,} denoted by $\Nbf_{\Kot,m,\Lambda}$
with metric $\gbf_{\Kot,m,\Lambda}$,
is 
the static spacetime with maximal compact spacelike slices, whose
lapse function $f_{\Kot, m,\Lambda}$ and 
Riemannian metric $g_{\Kot, m,\Lambda}$ 
on the compact spacelike slices 
$$
\Mcal_{\Kot, m,\Lambda} \simeq (r_\Kot^-,r_\Kot^+) \times S^2
$$ 
are defined by  
$$
(f_{\Kot,m,\Lambda}(r))^2 := 1 - {2m \over r} - {\Lambda \over 3} r^2, 
\quad
g_{\Kot,m,\Lambda} := {dr^2 \over (f_{\Kot,m,\Lambda}(r))^2} + r^2 \, g_{S^2}, \qquad r \in [r_\Kot^-,r_\Kot^+],  
$$
where $g_{S^2}$ denotes the canonical metric on the unit sphere $S^2$,  
$m$ is interpreted as the mass of the spacetime, and $r_\Kot^\pm = r_{\Kot^\pm,m,\Lambda}$
are the  two positive roots of 
 the cubic polynomial $r \mapsto r (f_{\Kot,m,\Lambda}(r))^2$.

These manifolds are also called {\em Schwarzschild-de~Sitter spacetimes} 
and 
provide us with a two-parameter
 family of static spacetimes with compact spacelike slices, which  
are locally (but not globally) conformally flat. 
Note that the horizon of a Kottler spacetime, denoted here by $\Hcal_{\Kot,m,\Lambda}$, consists of 
the two connected components   
$$
\Hcal^\pm_{\Kot,m,\Lambda} := \big\{ r = r^\pm_{\Kot,m,\Lambda} \big\}. 
$$   
We point out that the spacetimes $\Nbf_{\Kot,m,\Lambda}$ 
may be extended beyond their horizon: one component of $\Hcal_{\Kot,m,\Lambda}$
is an ``inner horizon'' connecting to an interior black hole region 
while the other component is a cosmological horizon connecting to a non-compact exterior domain of communication
(asymptotic to de~Sitter). The interior domain is, both mathematically and physically,  
the region of interest
and for instance, as $\Lambda \to 0$, converges to the outer communication domain of the Schwarzschild spacetime
dealt with in the classical black hole theorems.  

Finally, one more family of spacetimes are relevant in the present work, that is, the 
{\em de~Sitter spacetimes,} 
parametrized by their cosmological constant $\Lambda>0$. 
We denote by $\Nbf_{dS,\Lambda}$ one domain of communication of the de Sitter spacetime, 
whose spacelike slices
have the topology of a half-sphere $S^3$ and whose horizon $\Hcal_{dS,\Lambda}$
admits a single component diffeomorphic to the $2$-sphere $S^2$.

\section{Main results}

We are now in a position to state our ridigity results, under
the regularity condition that the level set achieving the maximum of
the lapse function is a regular surface.   

\vskip.25cm

\begin{theorem}[Uniqueness theorem for Kottler spacetime]  
\label{main}
The interior domain of the Kottler spa\-cetimes $\Nbf_{\Kot,m, \Lambda}$ 
parameterized by their mass $m>0$ and cosmological constant $\Lambda>0$ 
together with the domain of communication $\Nbf_{dS,\Lambda}$ of the de~Sitter spacetimes 
are, up to global isometries, the unique static spacetimes with maximal compact spacelike slices 
and regular maximal level set, 
satisfying Einstein's field equations with positive cosmological constant. 
\end{theorem}
   
\vskip.25cm

We emphasize that no restriction is assumed a~priori on 
the topology of the spacelike slices, and 
this topology is finally identified as part of  
the conclusion of the theorem, which also provides us with the metric. 
Hence, the above theorem is of interest in both general relativity and topology. 
A large literature is available on black hole uniqueness theorems, and 
we will not try to review it here but will only quote works that are most related to the 
present discussion.  

Classical works deal with the case ${\Lambda=0}$, and goes back to Israel \cite{Israel}, 
Hawking \cite{Hawking},  
and many others.  For more recent works, see Lindblom \cite{Lindblom} and 
Beig and Simon \cite{BeigSimon}.   
The class of (vacuum) spacetimes with negative cosmological constant ${\Lambda<0}$
was tackled only recently. (See \cite{LR1} for references.)  

In contrast with the above results 
and despite active research on the subject in the past twenty years, the class 
of spacetimes with positive cosmological constant is not amenable   
to the mathematical techniques developed 
in the existing literature.  
Our purpose in the present paper is to introduce a new approach 
which overcomes these (technical and conceptual) difficulties and 
to establish a 
uniqueness theorem for the case ${\Lambda>0}$. As we will show, 
we have to combine 
arguments from partial differential equations and differential geometry, 
and, most importantly, to work within a class of possibly singular foliations.

Our method of proof also applies in the Riemannian setting and
allows us to establish the validity 
of Besse conjecture \cite{Besse}. 
(See also the earlier works \cite{Kobayashi,Lafontaine} for special cases.)  

\vskip.25cm

\begin{theorem}[Besse conjecture in Riemannian geometry]
All compact three-manifolds $(M,g)$, on which there exists a non-trivial solution $f$ to the dual linearized curvature equation $L^*(f)=0$ with regular maximal level set, 
are given by the following list (up to isometries): 
\begin{itemize}

\item The sphere $S^3$ endowed with the canonical metric. In this case, one has
$f=\cos(d(.,x_0))$ where $d$ is the Riemannian distance to a point $x_0$,
and the kernel of $L^*$ has dimension $\dime \Ker(L^*)=4$.

\item A finite quotient of the product $S^1\times S^2$ endowed with the canonical product metric. In this case one has
$\dime \Ker(L^*)=2$. 

\item  A finite quotient of the twisted product $S^1\times S^2$ endowed with the metric $g=dx^2 + h^2(x) \, g_{S^2}$. 
These twisted products depend upon two real parameters and an integer parameter, and 
$\Ker(L^*)= h' \, \RR$.

\end{itemize}

\end{theorem} 

\vskip.25cm

\section{Elements of proof}

Let us indicate several key elements of our proof of Theorem~\ref{main}. 
We consider a 
static spacetime $\Nbf$ with maximal compact spacelike slices $\Mcal$
(and $W^{2,2}$ regularity) satisfying Einstein's field equations with positive cosmological constant $\Lambda>0$.   
Using the $(3+1)$-splitting, 
the Einstein equations on the $4$-dimensional spacetime are equivalent 
to a problem posed on the $3$-manifold $\Mcal$ with boundary, i.e.,  
to the partial differential equations (for the lapse function $f$ and metric $g$) 
$$
\nabla df - (\Delta f) \, g  - f \, \Rc = 0,
$$ 
with the additional constraint that the scalar curvature $R$ of $(\Mcal,g)$ coincides with) the
cosmological constant and, therefore,
is a constant; specifically, one has 
$
R = 2 \Lambda>0.
$
In the Einstein equations, the field of $1$-forms $df$ is the differential of $f$, while 
$\nabla$ denotes the covariant derivative in $(\Mcal,g)$, 
$\nabla	df$ the Hessian of $f$, 
$\Delta$ the Laplacian operator (normalized to have negative eigenvalues), 
and $\Rc$ the $3$-dimensional Ricci curvature, respectively. 
By taking the trace of the Einstein equations, we deduce that  
$$
\Delta f = - {R \over 2} f. 
$$
In other words, $f$ is an eigenfunction of the Laplace operator defined on the (unknown) 
Riemannian manifold $(\Mcal,g)$.
Our objective is to determine {\sl all triplets of solutions} $(\Mcal, g, f)$ satisfying 
the Einstein equations
and, in particular, to determine the topology of $\Mcal$.

From the lapse function associated with the natural $(3+1)$--foliation of the spacetimes
under consideration, we define a (possibly) degenerate $(2+1)$--foliation 
and investigate the topology and geometry of its leaves. 
It is convenient to introduce certain {\sl normalized} geometric invariants of this foliation,
which make sense globally on the manifold $\Mcal$, even at points where the gradient $\nabla f$ vanishes
(and the foliation possibly becomes degenerate.  
We also introduce the {\sl Hawking mass density,} 
defined from the Gauss curvature and mean-curvature of the $2$-slices, 
which again makes sense globally on the manifold, even at critical points. 
The Hawking mass density, used here, 
appears classically as an integrant in Hawking's original definition. 
Using the notion of Hawking mass density, we establish a pointwise version of Penrose inequality 
on the horizon, which allows us to identify a topological $2$-sphere within the connected components of the horizon.  
An ``optimal'' Kottler model with well-chosen ADM mass is introduced, which covers the region
limited by certain level sets of the lapse function.  
Finally, several maximum principle arguments are developed for Einstein's field equations of static spacetimes, 
which apply to the possibly degenerate $(2+1)$-foliation under consideration. 
For further details we refer to \cite{LR1,LR2}. 
 


\begin{thebibliography}{00} 
  
\bibitem{BeigSimon} \auth{Beig R. and Simon W.,}
On the uniqueness of static perfect-fluid solutions in general relativity,
Comm. Math. Phys. 144 (1992), 373--390.
 
\bibitem{Besse}  \auth{Besse A.L.,}
{\sl Einstein manifolds,} 
Springer Verlag, Ergebnisse der Mathematik, Vol. 10, 1987.    

\bibitem{Hawking} \auth{Hawking S.W.,}
Black holes in general relativity, 
Comm. Math. Phys. 25 (1972), 152--166. 

\bibitem{Israel} \auth{Israel W.,}
Event horizons in static vacuum spacetimes, 
Phys. Rev. 164 (1967), 1776--1779. 

\bibitem{Kobayashi} \auth{Kobayashi O.}, 
Scalar curvature of a metric with unit volume, 
Math. Ann. 279 (1987), 253-265. 

\bibitem{Kottler} \auth{Kottler F.,} 
Uber die physikalischen Grundlagen der Einsteinschen Gravitation theori,
Annalen der Physik 56 (1918), 401--402. 

\bibitem{Lafontaine} \auth{Lafontaine J.,}
Sur la g\'eom\'etrie d'une g\'en\'eralisation de l'\'equation diff\'erentielle d'Obata,
J. Math. Pures Appl. 62 (1983), 63--72. 

\bibitem{LR1} \auth{LeFloch P.G. and Rozoy L.,} 
A uniqueness theorem for Schwarzschild-de Sitter spacetime,  
submitted.

\bibitem{LR2} \auth{LeFloch P.G. and Rozoy L.,}  
in preparation.  
 
\bibitem{Lindblom} \auth{Lindblom L.,}
Static uniform-density stars must be spherical in general relativity,
J. Math. Phys. 29 (1988), 436--439.

\bibitem{Obata} \auth{Obata M.,} 
Certain conditions for a Riemanian manifold to be isometric with a sphere, 
J. Math. Soc. Japan 14 (1962), 333--340.  
 

\end{thebibliography}
\end{document}